\documentclass[preprint,preprintnumbers,amsmath,amssymb,nofootinbib]{revtex4}
\usepackage{xcolor}
\usepackage{amssymb}
\usepackage{amsmath}
\usepackage{cases}
\usepackage{epsfig,subfigure}
\usepackage{graphicx}
\usepackage{epstopdf}
\RequirePackage{graphicx}
\RequirePackage{mathptmx}      
\RequirePackage{flushend}
\RequirePackage[colorlinks,citecolor=blue,urlcolor=blue,linkcolor=blue]{hyperref}
\usepackage{tabularx}
\usepackage{bm}
\usepackage{multirow}

\def\bal#1\eal{\begin{align}#1\end{align}}
\newcommand\bseq{\begin{subequations}}
\newcommand\eseq{\end{subequations}}
\newcommand\beq{\begin{equation}}
\newcommand\eeq{\end{equation}}
\newcommand\beqn{\begin{eqnarray}}
\newcommand\eeqn{\end{eqnarray}}
\newcommand\nn{\nonumber}
\newcommand\fc{\frac}
\newcommand\lt{\left}
\newcommand\rt{\right}
\newcommand\pt{\partial}
\newcommand\tx{\text}

\allowdisplaybreaks

\begin{document}

\title{Boson Stars in Bumblebee Gravity and Their Gravitational Waveforms from Extreme-Mass-Ratio Inspirals}
\author{
Mao-Jiang Liu$^{a}$,
Long-Xing Huang$^{b,c,d}$,
Yong-Qiang Wang$^{b,c,d}$,
Ke Yang$^{a}$\footnote{keyang@swu.edu.cn, corresponding author}
}

\affiliation{
$^{a}$School of Physical Science and Technology, Southwest University, Chongqing 400715, China\\
$^{b}$Institute of Theoretical Physics and Research Center of Gravitation, Lanzhou University, Lanzhou 730000, China\\
$^{c}$Key Laboratory of Quantum Theory and Applications of MoE, Lanzhou University, Lanzhou 730000, China\\
$^{d}$Lanzhou Center for Theoretical Physics and Key Laboratory of Theoretical Physics of Gansu Province, Lanzhou University, Lanzhou 730000, China}

\begin{abstract}

We investigate the impact of Lorentz violation on the compactness of mini-boson stars and the resulting gravitational-wave signals from extreme-mass-ratio inspirals (EMRIs) within the framework of bumblebee gravity. Numerical solutions for static, spherically symmetric configurations reveal that a positive Lorentz-violating parameter $\ell$ suppresses repulsive pressure, thereby enhancing gravitational binding and yielding more compact boson stars. Conversely, a negative $\ell$ amplifies repulsive pressure and weakens gravitational binding, such that no static solutions exist beyond a critical negative value. These structural modifications imprint distinct features on EMRI dynamics, characterized by a monotonic decrease in both orbital eccentricity and radial range as $\ell$ gradually increases from negative to positive values. Unlike the intermittent bursts from grazing orbits that resemble black-hole signals, penetrating orbits that enter the boson-star core exhibit sustained, amplitude-modulated gravitational-wave signatures without quiet intervals. Their characteristic strain falls within the detectability range of LISA, providing a potential observable for constraining Lorentz violation.

\end{abstract}


\maketitle


\section{Introduction}

The direct detection of gravitational waves by ground-based observatories such as LIGO and Virgo has revolutionized our understanding of the Universe \cite{Abbott2016}. This breakthrough not only confirmed a cornerstone prediction of general relativity but also opened a new observational window onto the cosmos, yielding robust evidence for stellar-mass black holes and neutron stars through their mergers. Nevertheless, despite the profound success of Einstein's general relativity (GR) in describing gravitational phenomena, fundamental tensions persist, particularly with the principles of quantum mechanics and in explaining cosmic-scale observations related to dark matter and dark energy. These challenges continue to motivate the exploration of modified theories of gravity \cite{Clifton2012}.

Although empirical evidence strongly supports Lorentz symmetry as a fundamental property of nature, various approaches to quantum gravity suggest that this symmetry may be broken at the Planck scale. For instance, the emergence of nontrivial tensor vacuum expectation values in string theory \cite{Kostelecky1989a}, discrete spacetime structures in loop quantum gravity \cite{Alfaro2002}, and anisotropic scaling in Ho\v{r}ava-Lifshitz gravity \cite{Horava2009a} all point toward potential Lorentz violation. To systematically study the low-energy phenomenological consequences of such high-energy theories, effective field theory approaches are highly valuable. A particularly instructive framework in this context is the bumblebee gravity \cite{Kostelecky2004,Bailey2006}. As a key component of the gravitational sector of the Standard-Model Extension \cite{Kostelecky2004,Colladay1998}, this model provides a dynamic mechanism for spontaneous Lorentz symmetry breaking. It extends GR by incorporating a vector field—the bumblebee field—that acquires a nonzero vacuum expectation value (VEV), thereby naturally selecting a preferred spacetime direction while preserving the underlying diffeomorphism invariance. 

Interest in bumblebee models grew significantly after Casana et al.~derived an exact static, spherically symmetric solution \cite{Casana2018}. Since then, the theory has been widely employed to study a variety of black hole solutions \cite{Maluf2021,Jha2021,Ding2020a,Ding2021,Gullu2022,Xu2023,Ding2023,Ding2023a,Liu2025a,Liu2025b,Chen2025,Bailey2025,Li2025a,Li2025b}, along with their physical properties \cite{Kanzi2019,Liu2019,Oliveira2021,Liang2022,Wang2022,Wang2022a,Kuang2022,Chen2023,Liang2023,Zhang2023,Lin2023,Liu2023,Mai2023,Lambiase2023,Lambiase2024,Mai2024,An2024,Liu2024c,Pantig2025,Quan2025,Zhang2025,Yu2025,Liu2025c,Xu2025,Deng2025a,Li2025c,YuChih2025,Kumar2025}. Beyond black holes, wormhole geometries in this framework have been investigated in Refs.~\cite{Ovgun2019,Magalhaes2025,Ding2025}, and neutron star configurations in Refs.~\cite{Liu2024a,Ji2024}.

The existence of light scalar particles, motivated by particle-physics candidates for dark matter, raises the possibility that such fields could form compact, horizonless objects known as boson stars \cite{Khlopov1985}. These are self-gravitating condensates of bosonic fields, representing macroscopic quantum states in curved spacetime. The theoretical foundation for such objects was established by Kaup \cite{Kaup1968}, who discovered the first localized, regular solutions of the Einstein–Klein-Gordon system, and shortly thereafter by Ruffini and Bonazzola \cite{Ruffini1969}. This framework was later extended by Colpi et al.~\cite{Colpi1986}, who investigated a self-interacting scalar field with a quartic potential. Subsequent developments extending the scalar field to vector and spinor fields have enabled the construction of Proca stars \cite{Brito2016,SalazarLandea2016,Su2024,Zhang2025b} and Dirac stars \cite{Finster1999,Dzhunushaliev2019,Huang2024,Hao2024}, thereby broadening the family of horizonless compact objects. Being self-gravitating configurations of bosonic fields, these objects are compelling alternatives to black holes for interpreting gravitational-wave events \cite{CalderonBustillo2021,Evstafyeva2024}.

Boson stars have been studied in modified gravity theories as natural extensions of their GR counterparts, including in scalar-tensor theories \cite{Gunderson1993,Torres1997,Brihaye2016,Verbin2018,Evstafyeva2023}, Palatini $f(R)$ gravity \cite{Maso-Ferrando2021,Maso-Ferrando2024}, higher derivative models \cite{Baibhav2017,Wang2024a,Ma2025}, and teleparallel formulations \cite{Ilijic2020}. In this work, we investigate mini-boson stars—the simplest models governed by a scalar field with a quadratic potential—within the framework of bumblebee gravity. Our aim is to systematically analyze how Lorentz violation alters their compactness, stability, and potential observational signatures. 

This paper is organized as follows. Section \ref{Model} introduces the boson-star model in bumblebee gravity. Section \ref{Numerics} presents numerical solutions of the field equations and analyzes the effects of Lorentz violation. In Section \ref{EMRIs}, we study  extreme-mass-ratio inspirals (EMRIs) with a boson-star central object. Section \ref{Gravitational_Waves} investigates the impact of Lorentz violation on the resulting gravitational-wave emission. Finally, we present our conclusions.

\section{The model setup}\label{Model}

We begin with the action for bumblebee gravity \cite{Kostelecky2004}
\bal
S&=\int{}d^4x\sqrt{-g}\bigg[\frac{R}{2\kappa} - \frac{1}{4} B^{\alpha \beta} B_{\alpha \beta} + \xi B^\alpha B^\beta R_{\alpha \beta} - V\left(X\right)\bigg]+\int{}d^4x\sqrt{-g}\mathcal{L}_\tx{M},
\label{Main_Action}
\eal
where $\kappa \equiv 8\pi G$, $B_{\alpha\beta} \equiv \partial_{[\alpha} B_{\beta]}$ denotes the field strength of the bumblebee field, $\xi$ is the coupling constant between gravity and the bumblebee field, $V(X)$ is the potential of the bumblebee field with $X\equiv B^{\alpha} B_{\alpha} \pm b^2$, and $\mathcal{L}_\text{M}$ stands for the matter Lagrangian. Spontaneous Lorentz symmetry breaking is triggered when the bumblebee field acquires a nonzero VEV, $\langle B_\mu\rangle =b_\mu$, where $b_\mu$ is a spacetime-dependent vector field satisfying the constant norm condition \cite{Casana2018}, $b^\mu b_\mu = \pm b^2= \tx{const.}$ In the absence of a cosmological constant, the bumblebee potential $V$ is chosen such that both $V$ and $V_X\equiv dV/dX$ vanish at the VEV.

By setting the matter field to zero, $\mathcal{L}_\text{M}=0$, the Schwarzschild-like solution in this framework is derived as \cite{Casana2018}
\bal
ds^2&=-\lt(1-\fc{2M}{\sqrt{1+\ell}\,r}\rt)dt^2+\lt( \fc{1+\ell}{1-\fc{2M}{\sqrt{1+\ell}\,r}} \rt)dr^2+r^2\lt(d\theta^2+\sin^2\theta d\varphi^2 \rt),
\label{Metric_Vacuum}
\eal
where the mass $M$ of the central object is determined via the Wald formalism \cite{Wald1993}, which is crucial for obtaining the correct conserved charges and ensuring the thermodynamic consistency of black holes in this theory \cite{An2024,Li2025a}. The dimensionless Lorentz-violating parameter $\ell\equiv 2\kappa\xi b^2$ quantifies the strength of the Lorentz-violating effects.

Notably, taking $M \to 0$ in the metric \eqref{Metric_Vacuum} and performing the radial rescaling $r \to r / \sqrt{1+\ell}$, the background bumblebee VEV configuration gives a global vacuum solution
\begin{equation}
    ds^2 = -dt^2 + dr^2 + \lt(1-\Delta \rt) {r^2}\left( d\theta^2 + \sin^2\theta d\varphi^2 \right).
    \label{Vacuum_Solution}
\end{equation}
where $\Delta\equiv\frac{\ell}{1+\ell}$. This metric deviates from Minkowski spacetime for $\ell\neq0$, serving as a Lorentz-violating background. Specifically, it describes a global monopole and exhibits a solid angle deficit, which is singular at the origin $r \to 0$, as evidenced by the Ricci scalar $R = {2\ell}/{r^2}$ and the Kretschmann scalar $R^{\alpha\beta\gamma\delta}R_{\alpha\beta\gamma\delta} = {4\ell^2}/{r^4}$. In a complete field-theoretic description of a global monopole \cite{Barriola1989}, the point-like singularity is removed by a finite core of radius $\delta$, inside which the scalar field vanishes and the curvature remains finite.

Although the current model treats the origin as a point-like defect for simplicity, the subsequent analysis of boson star configurations remains physically self-consistent. This is formally justified by noting that the effective energy density of the background scales as $\rho_{\text{m}} \sim R \sim 1/r^2$. Consequently, the integrated mass contained within a radius $r$ is $M_{\text{m}}(r) = \int \rho_{\text{m}} \sqrt{-g} \, d^3x \propto r$, which implies that the mass contribution vanishes as $r \to 0$. As shown by Barriola and Vilenkin \cite{Barriola1989}, the gravitational potential $U \sim M_{\text{m}}(r)/r$ associated with this linear mass growth is constant throughout space. This results in a vanishing gravitational gradient, meaning the monopole background exerts no direct radial gravitational force on the surrounding matter for any $r > 0$. Therefore, the central singularity represents a purely topological vertex of a geometric deficit rather than a gravitational center that exerts a divergent radial pull.

We further stress that since spontaneous Lorentz symmetry breaking in bumblebee gravity is an infrared effect, Lorentz symmetry is expected to be restored in the ultraviolet. At high energies, the bumblebee field becomes fully dynamical rather than remaining frozen at the VEV. The approach to the origin, where curvature and energy density diverge, corresponds to the ultraviolet regime. In this regime, the bumblebee field can fluctuate away from the monopole-like VEV to regularize the geometry and smooth out the singularity. Meanwhile, the global topological feature persists in the infrared region to influence macroscopic dynamics. Therefore, the infrared-based analysis remains valid for describing the macroscopic properties and gravitational-wave signatures of boson stars.

The mini-boson star is modeled by a matter source composed of a massive $U(1)$ gauge scalar field without self-interaction. The corresponding Lagrangian density is given by
\bal
\mathcal{L}_\tx{M} = -\pt^{\alpha}{\Phi}^* \pt_{\alpha}\Phi - \mu^2 {\Phi}^* \Phi,
\label{Lagrangian_Scalar}
\eal
where $\mu$ denotes the boson mass.

Varying the action \eqref{Main_Action} with respect to the metric and the scalar field yields the gravitational field equation and the Klein-Gordon equation, respectively:
\bal
R_{\mu \nu }-\frac{1}{2} g_{\mu \nu }R&=\kappa \lt(T^\tx{(B)}_{\mu \nu }+T^{(\Phi)}_{\mu \nu }\rt), \label{Einstein_Eq}\\
\nabla^{\alpha }\nabla_{\alpha }\Phi &=\mu ^2 \Phi,\label{KG_Eq}
\eal
where the energy-momentum tensors for the bumblebee field and the scalar field read
\bal
T^\tx{(B)}_{\mu \nu }&=2 V_X B_{\mu } B_{\nu }-V g_{\mu \nu }-B_{\mu \alpha } B^{\alpha }{}_{\nu }-\frac{1}{4} g_{\mu \nu } B^{\alpha \beta } B_{\alpha \beta }\nn\\
&+\xi \bigg[ g_{\mu \nu } B^{\alpha } B^{\beta } R_{\alpha \beta }-2 B^{\alpha } B_{\mu } R_{\nu \alpha }-2 B^{\alpha } B_{\nu } R_{\alpha \mu }+\nabla _{\alpha }\nabla _{\mu }\left(B^{\alpha } B_{\nu }\right)\nn\\
&+\nabla _{\alpha }\nabla _{\nu }\left(B^{\alpha } B_{\mu }\right)-\nabla ^{\lambda }\nabla _{\lambda }\left(B_{\mu } B_{\nu }\right)-g_{\mu \nu } \nabla _{\alpha }\nabla _{\beta }\left(B^{\alpha } B^{\beta }\right)\bigg],\\
T^{(\Phi)}_{\mu \nu }&=\nabla_{\mu }\Phi^* \nabla_{\nu }\Phi +\nabla _{\mu }\Phi  \nabla_{\nu }\Phi^*-g_{\mu \nu } \big(\nabla^{\alpha }\Phi^* \nabla_{\alpha }\Phi + \mu ^2 \Phi^* \Phi \big).
\eal

We consider a static, spherically symmetric spacetime in the presence of Lorentz violation. The corresponding line element is assumed to take the form
\bal
ds^2\!=\!-n(r)\sigma^2(r)dt^2\!+\!\fc{1\!+\!\ell}{n(r)}dr^2\!+\!r^2\lt(d\theta^2\!+\!\sin^2\theta d\varphi^2 \rt),
\label{Metric_Boson}
\eal
where $\sigma(r)$ is an undetermined metric function, and $n(r)=1-\fc{2m(r)}{\sqrt{1+\ell}\,r}$, with $m(r)$ being the mass function that characterizes the mass distribution of the boson star, i.e., the total mass enclosed within a coordinate radius $r$. At spatial infinity ($r \to \infty$), the complex scalar field vanishes asymptotically. Matching the exterior vacuum solution \eqref{Metric_Vacuum} requires the asymptotic conditions $\sigma(r \to \infty) \to 1$ and $m(r \to \infty) \to M$, where $M$ denotes the total mass of the boson star. 

Here, we note that in the presence of spontaneous Lorentz symmetry breaking, the conventional definitions of mass encounter fundamental difficulties. Specifically, the non-flat asymptotic behavior induced by the solid-angle deficit ($\ell \neq 0$) renders the Arnowitt-Deser-Misner (ADM) mass integral ill-defined. Furthermore, beyond the violation of the first law of thermodynamics, the Komar mass is inadequate in this framework as it fails to account for the additional Noether contributions arising from non-minimal interactions between the bumblebee field and the curvature sector \cite{An2024,Li2025a}. 

To properly define a conserved total mass $M$, we adopt the Wald formalism \cite{Wald1993}, where the mass variation is determined via the Hamiltonian boundary integral at spatial infinity, i.e., $\delta M = \int_{S_{\infty}} (\delta \bm{Q}_{\xi} - i_{\xi} \bm{\Theta})$. For the system under consideration, the minimally coupled scalar field yields no contribution to the Noether charge 2-form $\bm{Q}_{\xi}$. However, the scalar sector does contribute to the symplectic potential $\bm{\Theta}$ through terms involving the field and its gradients, specifically $\bm{\Theta} \supset- \nabla^\mu \Phi^* \delta \Phi - \nabla^\mu \Phi \delta \Phi^*$. Crucially, since the boson star is a localized bound-state configuration, the scalar field and its radial derivatives exhibit exponential decay as $r \to \infty$. This rapid fall-off ensures that the scalar field's contribution to the surface integral vanishes identically at the spatial boundary.

 Consequently, the functional form of the Wald mass variation for a boson star reduces to the same geometric expression as that of the vacuum Schwarzschild-like configuration \cite{An2024}, yielding $\delta \bm{Q}_{\xi} - i_{\xi} \bm{\Theta} = -\sqrt{1+\ell}r\sigma\delta n/{\kappa}$. With the metric \eqref{Metric_Boson} and the associated asymptotic conditions, the conserved total mass $M$ corresponds precisely to the asymptotic limit of the mass function, $M= m(\infty)$. Consequently, we compute the total mass $M$ by directly extracting the value of the mass function $m(r)$ at the asymptotic boundary from the numerical solution.

For stationary mini-boson star configurations, we adopt the standard harmonic ansatz
\bal
\Phi(t,r)=\phi(r)e^{i \omega t},
\label{Harmonic_ansatz}
\eal
where $\phi(r)$ is a real radial function and $\omega$ is a real constant representing the oscillation frequency of the scalar field. Since the scalar Lagrangian \eqref{Lagrangian_Scalar} is globally $U(1)$ invariant, the associated Noether current takes the form
\bal
J^\alpha=i g^{\alpha\beta}\lt(\Phi^* \pt_\beta \Phi -\Phi \pt_\beta \Phi^* \rt).
\eal
The corresponding conserved charge $N$ is obtained by integrating the time component of the current over a space-like hypersurface $\Sigma$, yielding
\bal
N = \int_\Sigma \sqrt{-g} J^t dr d\theta d\varphi = 8\pi \omega \sqrt{1+\ell}\int_\Sigma \fc{r^2 \phi^2}{n\sigma}  dr.
\label{Charge}
\eal
This charge corresponds to the total number of bosons in the star.

Following the study \cite{Casana2018}, we adopt a purely radial, spacelike vacuum configuration for the bumblebee field, $b_\mu = (0, b\sqrt{(1+\ell)/n(r)}, 0, 0)$. The configuration is chosen to be consistent with the static and spherically symmetric properties of the spacetime while strictly adhering to the constant norm condition $b^\mu b_\mu = b^2$. Furthermore, this standard ansatz simplifies the resulting field equations by ensuring that the bumblebee field strength vanishes identically. Substituting this configuration, along with the metric and scalar harmonic ansatzes, into the field equations \eqref{Einstein_Eq} and \eqref{KG_Eq} yields the following reduced system of coupled ordinary differential equations (ODEs):
\bal
m' &= 4 \pi G \sqrt{1+\ell}r^2\lt[\lt(\mu^2 +\fc{\omega^2}{n \sigma^2} \rt)\phi^2+\fc{n \phi'^2}{1+\ell}\rt],\\
o' &= \fc{8 \pi G r \sigma}{2+3 \ell} \lt[\fc{2+\ell}{1+\ell}\phi'^2-\lt(\mu^2 \ell-\fc{(2+3 \ell)\omega ^2}{n \sigma^2}\rt)\frac{\phi^2}{n}\rt],\\
\phi'' &=(1+\ell)\lt(\mu ^2-\frac{\omega ^2}{n \sigma^2}\rt) \fc{\phi }{n}-\lt(\fc{2}{r}+\fc{n'}{n}+\fc{\sigma'}{\sigma}\rt)\phi',
\eal
where the prime denotes differentiation with respect to the radial coordinate $r$. 

To solve these ODEs and construct physically viable boson star solutions, appropriate boundary conditions are essential to ensure regularity and consistency. At the origin ($r \to 0$), regularity of $n(r)$ together with a small-$r$ analysis of the ODEs yield
\bal
m(0) = 0, \quad \sigma(0) = \sigma_0, \quad \phi'(0) = 0
\eal
At spatial infinity ($r \to \infty$), matching the metric \eqref{Metric_Boson} to the Schwarzschild‑like exterior solution \eqref{Metric_Vacuum} and requiring the scalar field to vanish asymptotically give
\bal
m(\infty) = M, \quad \sigma(\infty) = 1, \quad \phi(\infty) = 0.
\eal
The constants $\sigma_0$ and the boson star mass $M$ are then determined by solving the ODE system.

\section{Numerical Results}\label{Numerics}

To facilitate numerical computation, we adopt the following scaling transformations to render the variables dimensionless, i.e.,
\bal
r \to \frac{r}{\mu}, 
 \quad \omega \to \mu {\omega},
 \quad m \to \fc{m}{\mu},
 \quad \phi \to \frac{\phi}{\sqrt{4\pi G}}.
 \eal
This scaling is technically equivalent to adopting the units $4\pi G = 1$ and $\mu = 1$ in the numerical implementation. To enable numerical treatment over the entire spatial domain, we employ a compact radial coordinate $x = \frac{r}{1+r}$, which maps the semi-infinite interval $r\in [0, \infty)$ to the unit interval $x\in [0, 1]$. To solve the nonlinear ODEs, we employ the finite element method with 1000 grid points over the integration domain $[0, 1]$. The Newton-Raphson method is used for iteration, and to ensure numerical accuracy, we enforce a relative error below $10^{-5}$. Consequently, the boson star solutions are fully characterized by the parameters $\omega$ and $\ell$.

\begin{figure}[t]
\begin{center}
\includegraphics[width=7cm,height=5cm]{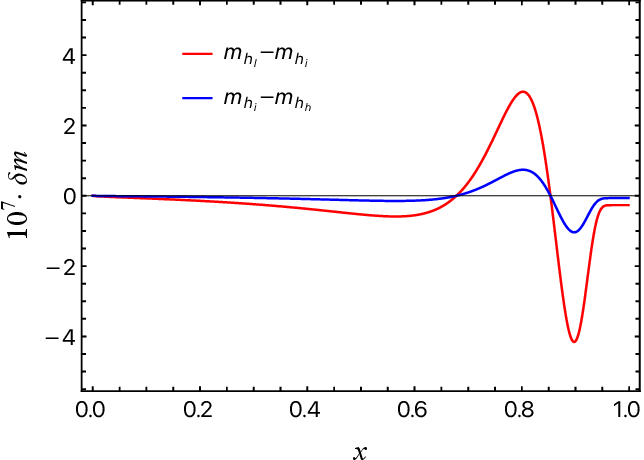}
\caption{Convergence test for the mass function $m(x)$. The red and blue curves represent the residuals between coarse-intermediate $(m_{h_l}-m_{h_i})$ and intermediate-high $(m_{h_i}-m_{h_h})$ resolutions, respectively.}
\label{Convergence_Mass}
\end{center}
\end{figure}

To verify the accuracy and robustness of the numerical solutions for the boson star configurations, we perform a self-convergence test for the mass function $m(x)$. We compare the results using three distinct grid resolutions by varying the number of grid points $N$ over the integration  domain $[0, 1]$. We define the grid spacing as $h = 1/N$ and consider three cases: coarse ($N_l = 1000, h_l = 10^{-3}$), intermediate ($N_i = 2000, h_i = 0.5 \times 10^{-3}$), and fine ($N_h = 4000, h_h = 0.25 \times 10^{-3}$). As illustrated in Fig.~\ref{Convergence_Mass}, the residuals between successive resolutions ($m_{h_l}-m_{h_i}$ and $m_{h_i}-m_{h_h}$) decrease systematically. The calculated convergence factor, $Q = |m_{h_l} - m_{h_i}| / |m_{h_i} - m_{h_h}| \approx 4$ is approximately 4. This aligns with the  theoretical value of 4 for the finite element scheme, demonstrating that the numerical error is well-controlled and our results presented using 1000 grid points are highly reliable.

\begin{figure*}[t]
\begin{center}
\subfigure[~$\phi(r)$]  {\label{Fig_phi_r}
\includegraphics[width=7cm,height=5cm]{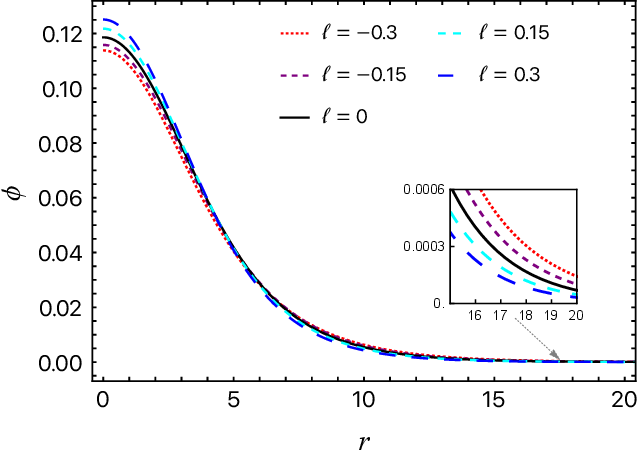}}\qquad
\subfigure[~$m(r)$]  {\label{Fig_m_r}
\includegraphics[width=7cm,height=5cm]{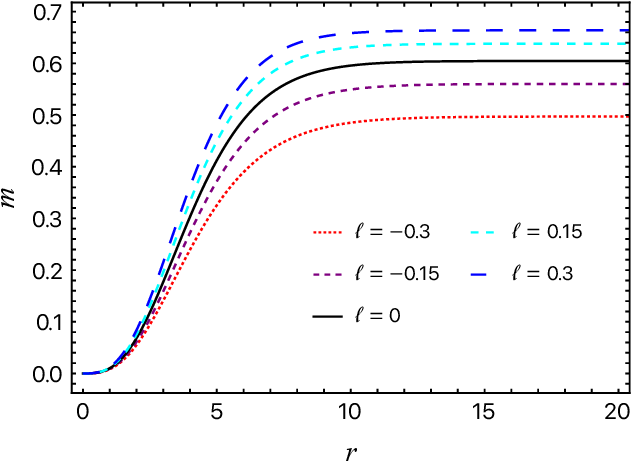}}
\subfigure[~$n(r)$]  {\label{Fig_n_r}
\includegraphics[width=7cm,height=5cm]{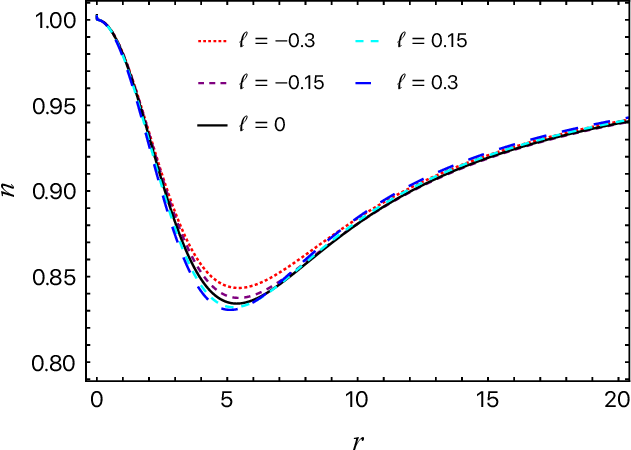}}\qquad
\subfigure[~$\sigma(r)$]  {\label{Fig_o_r}
\includegraphics[width=7cm,height=5cm]{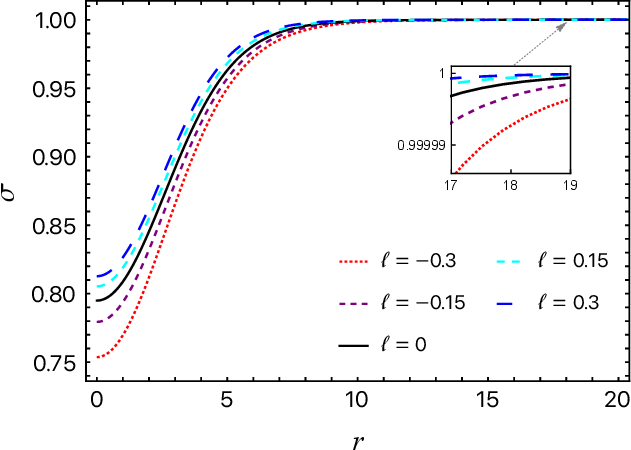}}
\caption{$\phi(r)$, $m(r)$, $n(r)$ and $\sigma(r)$ for different values of the Lorentz-violating parameter $\ell$, with $\omega=0.9$. Note that the case $\ell=0$ corresponds to the GR limit, whereas $\ell=\pm0.1, \pm0.3$ presented here exceed current experimental bounds and are chosen for illustrative clarity to demonstrate qualitative trends.}
\label{Fig_phi_m_sigma}
\end{center}
\end{figure*}

\begin{figure*}[t]
\begin{center}
\subfigure[~$M(\omega)$]  {\label{Fig_M_omega}
\includegraphics[width=7cm,height=5cm]{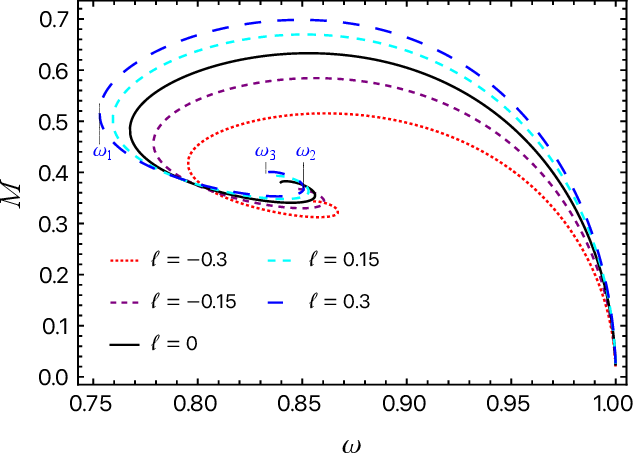}}\qquad
\subfigure[~$M(\ell)$]  {\label{Fig_M_l}
\includegraphics[width=7cm,height=5cm]{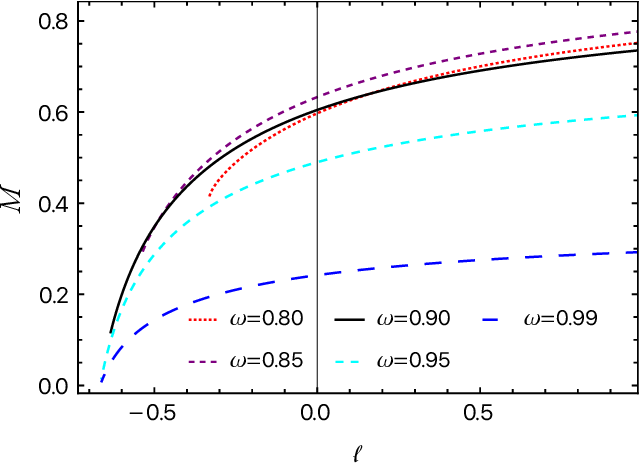}}
\caption{Mass $M$ plotted against the frequency $\omega$ and the Lorentz-violating parameter $\ell$, respectively. Panel (a) highlights the characteristic spiral structure, where backbending points of $\omega$ act as boundaries between distinct branches. On the representative $\ell=0.3$ curve, the first three backbending frequencies ($\omega_1, \omega_2, \omega_3$) are explicitly labeled.}
\label{Fig_MN_omega}
\end{center}
\end{figure*}

\begin{figure*}[t]
\begin{center}
\subfigure[~$N(\omega)$]  {\label{Fig_N_omega}
\includegraphics[width=7cm,height=5cm]{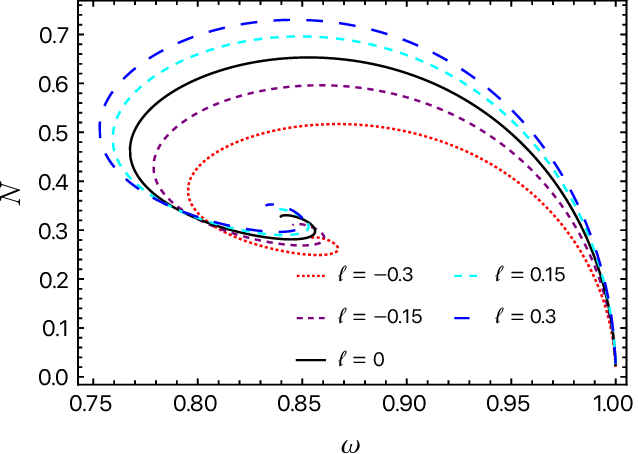}}\qquad
\subfigure[~$N(\ell)$]  {\label{Fig_N_l}
\includegraphics[width=7cm,height=5cm]{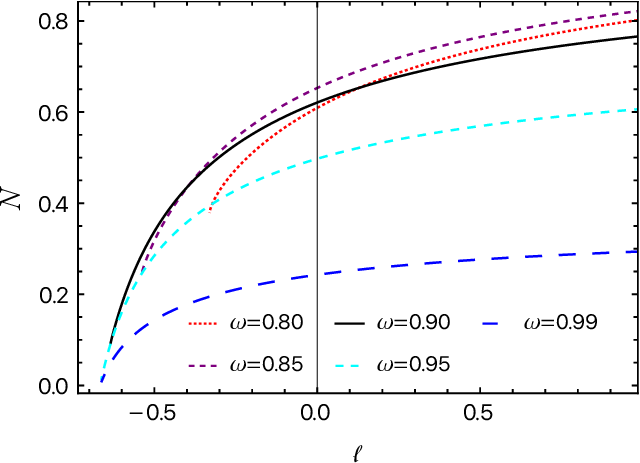}}
\caption{Particle number $N$ plotted against the frequency $\omega$ and the Lorentz-violating parameter $\ell$, respectively.}
\label{Fig_Mass}
\end{center}
\end{figure*}

\begin{figure*}[t]
\begin{center}
\subfigure[~$C(\omega)$]  {\label{Fig_C_omega}
\includegraphics[width=7cm,height=5cm]{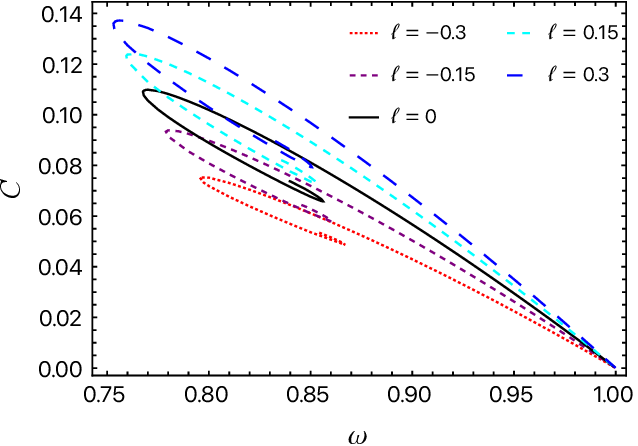}}\qquad
\subfigure[~$C(\ell)$]  {\label{Fig_C_l}
\includegraphics[width=7cm,height=5cm]{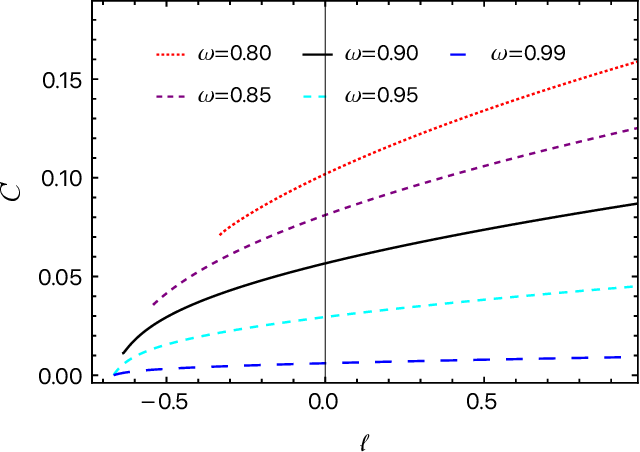}}
\caption{Compactness $C$ plotted against the frequency $\omega$ and the Lorentz-violating parameter $\ell$, respectively.}
\label{Fig_M_R_C_omega}
\end{center}
\end{figure*}

The numerical solutions for the scalar field $\phi(r)$, the mass function $m(r)$, and the metric functions $n(r)$ and $\sigma(r)$ are shown in Fig.~\ref{Fig_phi_m_sigma} for different values of the Lorentz-violating parameter $\ell$, where $\ell=0$ corresponds to the GR limit. We note that Lorentz-violating effects are subject to stringent experimental constraints \cite{Casana2018,Lambiase2024,Yang2023,Duan2024,Junior2024,Pantig2025,Kumar2025}. For instance, solar system tests restrict the parameter to $|\ell| < 10^{-10}$ \cite{Casana2018}, and consequently, the values $\ell=\pm 0.1, \pm0.3$ adopted in the figures lie outside the physically allowed regime. These parameters are chosen for illustrative clarity, as realistic values would be visually indistinguishable from GR in the plots. On the other hand, we aim to theoretically explore the potential impact of significant Lorentz-violating effects on the physical properties of boson stars. As illustrated later in Figs.~\ref{Fig_M_l}, \ref{Fig_N_l} and \ref{Fig_C_l}, the dependence of the total mass $M$, particle number $N$ and compactness $C$ on the parameter $\ell$ is characterized by smooth and monotonic behavior. This suggests that the qualitative trends identified within this amplified parameter range remain representative of the static background configurations within the realistic, experimentally constrained regime. For the EMRI dynamics and gravitational-wave signatures, however, we restrict our quantitative discussion to the specific parameter values investigated ($\ell = 0, \pm 0.01$). These results effectively demonstrate the qualitative impact of Lorentz violation, though such behaviors in a broader parameter space would necessitate further exhaustive investigation.

\begin{table*}[h]
\centering
\begin{tabular}{|c|c|c|c|c|c|c|c|c|}
     \hline
   $\ell$ & $M_\tx{max}$  &  $\omega\lt(M_\tx{max}\rt)$ &  $\omega_1$   & $\omega_2$  & $\omega_3$  & $\Delta \omega_\tx{B1}$ & $\Delta \omega_\tx{B2}$ & $\Delta \omega_\tx{B3}$  \\
     \hline
      -0.3   & 0.6322 & 0.8612  & 0.7954 & 0.8670  & 0.8543 & 0.2046 &  0.0716 &  0.0127  \\
       \hline
     -0.15 & 0.6329 & 0.8562  & 0.7789 & 0.8604  & 0.8455 & 0.2211 & 0.0815 & 0.0149  \\
       \hline
      0  & 0.6330 & 0.8530  & 0.7677 & 0.8561  & 0.8398 & 0.2323 & 0.0884 & 0.0163  \\
       \hline
      0.15  & 0.6329 & 0.8508  & 0.7594 & 0.8530  & 0.8356 & 0.2406 & 0.0936 & 0.0174  \\
       \hline
       0.3   & 0.6328 & 0.8492 &  0.7531 & 0.8506  & 0.8326  & 0.2469 &  0.0975 & 0.0180 \\ 
       \hline
\end{tabular}
\caption{The maximum mass $M_\tx{max}$ and its corresponding frequency $\omega(M_\tx{max})$, the frequencies at the first (lowest), second, and third backbending points $\omega_1$, $\omega_2$, $\omega_3$, and the frequency range of the $i$-th branch $\Delta \omega_{\tx{B}i}$.}
\label{Tab_M_omega}
\end{table*}

The radial profile $\phi(r)$ decreases monotonically from its central value $\phi(0)$ toward the asymptotic vacuum. As the Lorentz-violating parameter $\ell$ increases from negative to positive values, the central amplitude $\phi(0)$ rises, and the field decays more rapidly, indicating enhanced radial confinement. The mass function $m(r)$ increases monotonically from the origin and approaches its asymptotic value, the total mass $M$. As shown in Fig. \ref{Fig_M_l}, the mass $M$ increases monotonically as $\ell$ increases from negative to positive values, passing smoothly through the GR limit ($\ell=0$). In particular, for sufficiently negative values of $\ell$, the system admits no static solutions. The corresponding mass curves terminate at $\ell = (-0.331, -0.540, -0.634, -0.663, -0.663)$ for frequencies $\omega = (0.8, 0.85, 0.9, 0.95, 0.99)$.

As illustrated in Fig.~\ref{Fig_M_omega}, the $M(\omega)$ curves exhibit the characteristic spiral form, starting from the point $M=0$ at $\omega=1$, irrespective of the magnitude of the Lorentz-violating parameter. These curves are segmented into distinct branches by backbending points occurring at specific frequencies, which are denoted as $\omega_i$ in the figure. As summarized in Tab.~\ref{Tab_M_omega}, our numerical results show that the allowed frequency ranges of all three branches broaden as the Lorentz-violating parameter $\ell$ increases from negative to positive values. This is accompanied by a decrease in the characteristic frequencies: the minima of the first and third branches ($\omega_{\text{1st}}$, $\omega_{\text{3rd}}$) and the maximum of the second branch ($\omega_{\text{2nd}}$). In summary, a positive $\ell$ enhances the boson-star mass and broadens the solution space, while a negative $\ell$ reduces the mass and narrows the solution space.

The dependence of the particle number $N$ on the frequency $\omega$ and on the Lorentz-violating parameter $\ell$ is shown in Fig.~\ref{Fig_Mass}. These figures exhibit trends similar to those of the mass. Specifically, the particle number increases monotonically as $\ell$ varies continuously from negative to positive values. 

\begin{figure}[t]
\begin{center}
\includegraphics[width=7cm,height=5cm]{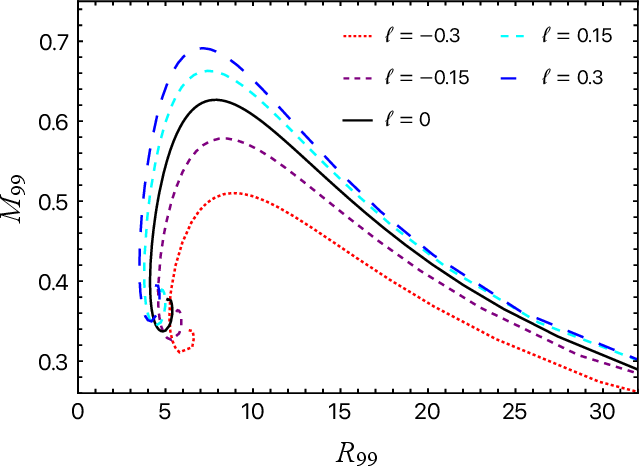}
\caption{$M_{99}$-$R_{99}$ curve as a function of frequency $\omega$ for different values of the Lorentz-violating parameter $\ell$.}
\label{Fig_M99_R99}
\end{center}
\end{figure}
 
The presence of Lorentz violation significantly modulates the compactness of boson stars, defined as $C \equiv M_{99}/R_{99}$, where $R_{99}$ denotes the radius enclosing 99\% of the total mass, and $M_{99}$ is the mass enclosed within that radius. Fig.~\ref{Fig_C_omega} illustrates the relationship between $C$ and the frequency $\omega$, while Fig.~\ref{Fig_C_l} explicitly demonstrates the dependence of $C$ on the Lorentz-violating parameter $\ell$. It is evident that $C$ increases monotonically as $\ell$ varies from negative to positive values. This trend implies that a positive $\ell$ suppresses the spatial dispersion of the scalar field, favoring a more concentrated configuration. Conversely, a negative $\ell$ enhances the outward delocalization of the scalar field, thereby weakening the gravitational binding of the system.

As illustrated in Fig.~\ref{Fig_C_l}, positive values of the Lorentz-violating parameter $\ell$ significantly enhance the compactness $C$. Specifically, for $\omega=0.8$, the compactness attains $C \approx 0.16$, a value that approaches the typical regime of neutron stars ($C \approx 0.2$). Nevertheless, the enhanced compactness remains markedly lower than the event horizon limit of black holes ($C_{\mathrm{BH}} \approx 0.5$) or the light ring limit of the ultra-compact thin-shell boson stars ($C > 1/3$) \cite{Cardoso2019,Cardoso2022,Collodel2022,Mazur2023}. This implies that although Lorentz violation promotes a more concentrated configuration, its impact on increasing compactness is relatively moderate.

The decrease in compactness for negative $\ell$ reflects a repulsive effect, providing a point of reference to the well-known repulsive quartic self-interaction ($\lambda|\Phi|^4, \lambda > 0$). However, these two mechanisms lead to opposite outcomes. Unlike the quartic model, where repulsion increases mass capacity and peak compactness \cite{Colpi1986}, the repulsion in Bumblebee gravity weakens gravitational binding and lowers the star's concentration. This difference arises because Lorentz violation acts as a geometric modification of the gravitational sector, rather than a scalar field self-coupling.

The corresponding mass-radius relations, displayed in Fig.~\ref{Fig_M99_R99}, exhibit the characteristic inspiral behavior. For a fixed effective radius $R_{99}$, the mass $M_{99}$ increases as $\ell$ increases from negative to positive values. This is because a positive $\ell$ suppresses diffusion, whereas a negative $\ell$ enhances it. The mass and effective radius in SI units are derived from the dimensionless quantities $M_{99}$ and $R_{99}$ via $M_\tx{SI}=\fc{\hbar c M_{99}}{G \mu}$ and $R_\tx{SI}=\fc{\hbar R_{99}}{\mu c}$, where the factor $\fc{\hbar}{\mu c}$ in $R_\tx{SI}$ corresponds to the reduced Compton wavelength of the boson. For instance, with ultra-light bosons with mass $\mu \sim 10^{-16}\,$eV, the boson star scales as $M_\tx{SI}\sim 10^6 M_\odot$ and $R_\tx{SI}\sim 10^7\,$km.

\begin{figure}[t]
\begin{center}
\includegraphics[width=7cm,height=5cm]{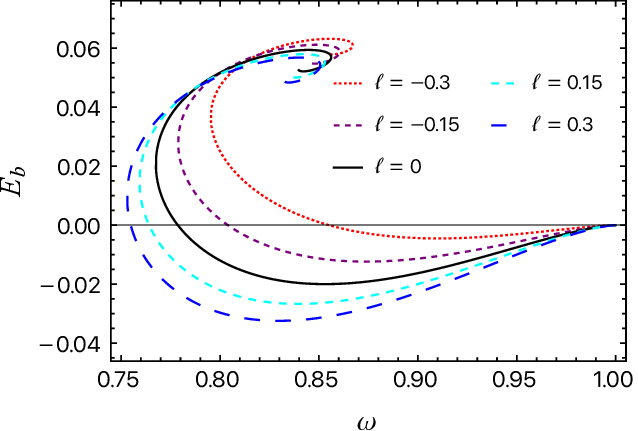}
\caption{Binding energy $E_b$ plotted against the frequency $\omega$ for different values of the Lorentz-violating parameter $\ell$.}
\label{Fig_Binding_Energy}
\end{center}
\end{figure}

The energetic stability of these boson stars can be analyzed through the binding energy $E_b = M - \mu N$, where $E_b < 0$ indicates a gravitationally bound configuration. As illustrated in Fig.~\ref{Fig_Binding_Energy}, the binding energy decreases monotonically as the Lorentz-violating parameter $\ell$ increases from negative to positive values, thus extending the frequency range over which bound states exist. For positive $\ell$, the suppression of diffusion provides an effective attraction that enhances cohesion. In contrast, for negative $\ell$, the promoted diffusion introduces an effective repulsion, which tends to destabilize the star. Specifically, the zero-binding frequencies are identified as $\omega(E_b=0) = (0.8549, 0.8039, 0.7786, 0.7641, 0.7548)$ for $\ell = (-0.3, -0.15, 0, 0.15, 0.3)$, respectively.

While these values mark the energetic stability thresholds, dynamical stability against radial perturbations represents a more stringent criterion for physical viability \cite{Liebling2023}. In this regard, the dynamically stable branch along the $M(\omega)$ curve extends from the vacuum limit $\omega=1$ down to the first maximum mass $M_{\text{max}}$, as illustrated in Fig.~\ref{Fig_M_omega}. A comparison between these zero-binding frequencies and the dynamical stability limits $\omega(M_{\text{max}})$ listed in Tab.~\ref{Tab_M_omega} reveals that $\omega(M_{\text{max}}) > \omega(E_b=0)$. This indicates that as the frequency decreases from the vacuum limit $\omega=1$, boson stars encounter the onset of dynamical instability before their binding energy vanishes. Furthermore, as $\ell$ increases from $-0.3$ to $0.3$, the threshold $\omega(E_b=0)$ decreases more sharply than $\omega(M_{\text{max}})$, indicating that the energetic stability boundary is significantly more sensitive to the Lorentz-violating parameter $\ell$ than the dynamical boundary.

\begin{figure}[t]
\begin{center}
\includegraphics[width=7cm,height=5cm]{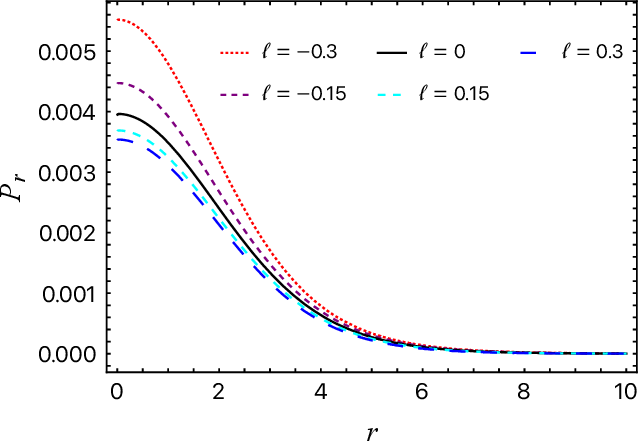}
\caption{Radial pressure $P_r(r)$ for different values of the Lorentz-violating parameter $\ell$.}
\label{Fig_Radial_Pressure}
\end{center}
\end{figure}

The observed Lorentz-violating effect can be intuitively understood from its impact on the radial pressure $P_r$, which provides the hydrodynamic support against gravity:
\begin{equation}
P_r = T^{(\Phi)}{}^r{}_r = \frac{n}{1+\ell}\phi'^2 + \frac{\omega^2}{n\sigma^2}\phi^2 - \mu^2 \phi^2.
\label{Eq_Pr}
\end{equation}
Here, the first term stems from the spatial gradient and contributes a positive pressure, the second term arises from the time-kinetic energy and also supplies a repulsive effect, while the third term, coming from the mass, yields a negative binding pressure. As shown in Fig.~\ref{Fig_Radial_Pressure}, the radial pressure is suppressed for $\ell>0$ relative to the GR case $\ell=0$, which weakens the diffusive capacity of the bosonic matter. Conversely, the radial pressure is enhanced for $\ell<0$, thereby strengthening their diffusive capacity. In particular, once $\ell$ falls below a critical negative threshold, the excessive diffusion prevents the boson star from sustaining a static equilibrium, so that static solutions no longer exist, as illustrated in Fig.~\ref{Fig_M_l}.

\section{Orbital motion of test particles and EMRIs}\label{EMRIs}

Our analysis demonstrates that Lorentz violation not only shifts the boundaries of the viable parameter space for boson-star solutions but also appreciably modifies their internal structure. The resulting modifications produce potentially detectable imprints on astrophysical observables, such as gravitational-wave signals, lensing profiles, etc. EMRIs provide a premier laboratory for probing strong-field gravity, featuring a stellar-mass compact object spiraling into a supermassive central body. In this section, we analyze the impact of Lorentz violation on an EMRI system consisting of a central boson star of $10^6 M_\odot$ and a stellar‑mass compact object of $10 M_\odot$, which can be modeled as a test particle.

The geodesic motion of a test particle is governed by the Lagrangian
\beq
\mathcal{L}=-\fc{1}{2}g_{\alpha\beta}\dot{x}^\alpha\dot{x}^\beta,
\label{Lagrangian_Abst}
\eeq
where the overdot denotes differentiation with respect to the affine parameter $\lambda$. For timelike geodesics of massive particles, the Lagrangian satisfies $\mathcal{L} = \frac{1}{2}$.

Given the spherical symmetry of the spacetime, without loss of generality, here we focus on equatorial geodesics with $\theta = \pi/2$. Consequently, we obtain the equation from the Lagrangian \eqref{Lagrangian_Abst} as 
\bal
n(r)\sigma^2(r) \left( \frac{dt}{d\lambda} \right)^2 - \fc{1+\ell}{n(r)}\left( \frac{dr}{d\lambda} \right)^2 - r^2 \left( \frac{d\varphi}{d\lambda} \right)^2 = 1.
\label{Lagrangian_Expl}
\eal

\begin{figure*}[t]
\begin{center}
\subfigure[~$(L=3, E=0.985)$]  {\label{Fig_Potential_L3E0985}
\includegraphics[width=7cm,height=5cm]{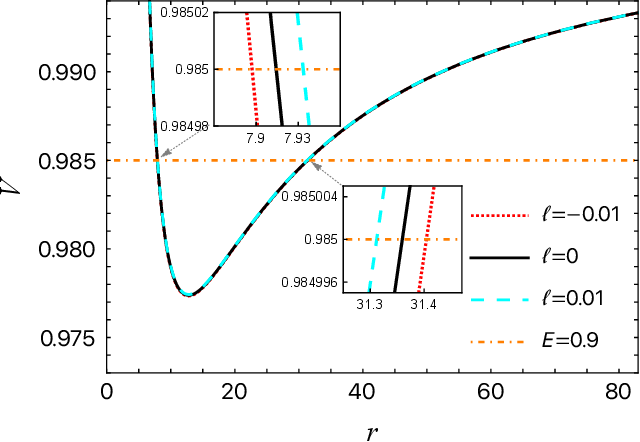}}\qquad
\subfigure[~$(L=0.5, E=0.9)$]  {\label{Fig_Potential_L05E09}
\includegraphics[width=7cm,height=5cm]{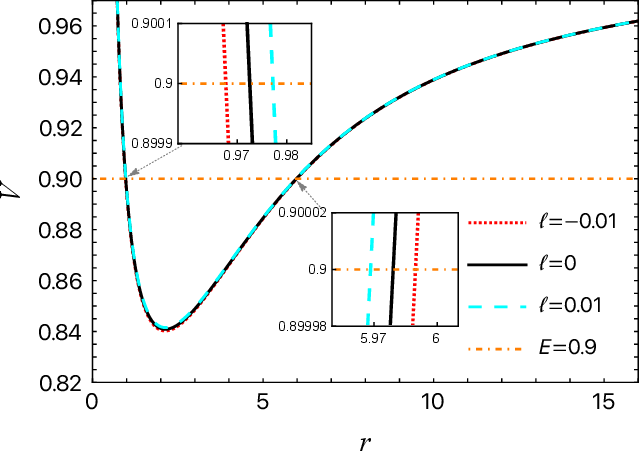}}
\caption{Effective potential $V(r)$ for different values of the Lorentz-violating parameter $\ell$, corresponding to $(L=3, E=0.985)$ and $(L=0.5, E=0.9)$. Note that the case $\ell=0$ corresponds to the GR limit, whereas $\ell=\pm 0.01$ are chosen to magnify the specific dynamical effects for illustrative purposes, as they exceed current experimental bounds.}
\label{Fig_Potentials} 
\end{center}
\end{figure*}

Exploiting the two conserved quantities in static, spherically symmetric spacetimes, i.e., the specific energy $E = \frac{\partial \mathcal{L}}{\partial \dot{t}} = n(r)\sigma^2(r) \dot{t}$ and the specific angular momentum $L = -\frac{\partial \mathcal{L}}{\partial \dot{\varphi}} = r^2 \dot{\varphi}$, one has
\bal
\dot{r}^2 = \fc{E^2}{(1+\ell) \sigma(r)^2}-\fc{n(r)}{1+\ell}\lt(1+\fc{L^2}{r^2}\rt).
\label{Orbital_Eq}
\eal
The effective potential is obtained by setting $\dot{r}^2=0$, yielding
\bal
V_\tx{eff} = \sigma(r) \sqrt{n(r) \lt(1+\fc{L^2}{r^2}\rt)}.
\eal

Analysis of the effective potential provides valuable insights into particle motion within the boson star background. Fig.~\ref{Fig_Potentials} displays the effective potential $V(r)$ for different values of the Lorentz-violating parameter $\ell$, illustrated for two distinct sets of specific angular momentum and energy, $(L, E)$. The parameter set $(L=3, E=0.985)$ corresponds to the orbits farther from the center of the boson star, whereas $(L=0.5, E=0.9)$ corresponds to orbits that probe deeper into its central region.

The radial velocity $v_r \equiv \dot{r}$ and the angular velocity $\dot{\phi}$ can be obtained from Eq.~(\ref{Orbital_Eq}) and $L = r^2 \dot{\varphi}$. By integrating these velocities, one can plot the trajectories of the test particles. However, for numerical implementation, the second-order approach is particularly advantageous as it naturally handles the sign transitions of the radial velocity at the orbital turning points without numerical discontinuities. To integrate the particle trajectories numerically, we transform the second-order geodesic equations into a coupled system of first-order ODEs as
\begin{equation}
    \begin{bmatrix} 
    \dot{t} \\ \dot{r} \\ \dot{\varphi} \\ \dot{v_r} 
    \end{bmatrix} =
    \begin{bmatrix} 
    \fc{E}{n\sigma^2} \\ 
    v_r \\ 
    \fc{L}{r^2} \\ 
    \fc{n'}{2n} v_r^2 + \frac{nr}{1+\ell} \dot{\varphi}^2 -\fc{n^2\sigma^2}{2(1+\ell)}\lt(\fc{n'}{n}+\fc{2\sigma'}{\sigma} \rt)\dot{t}^2
    \end{bmatrix}.
    \label{Eq_ODE_System}
\end{equation}
This system is solved using a fourth-order Runge-Kutta (RK4) method.

\begin{figure}[t]
\begin{center}
\includegraphics[width=7cm,height=5cm]{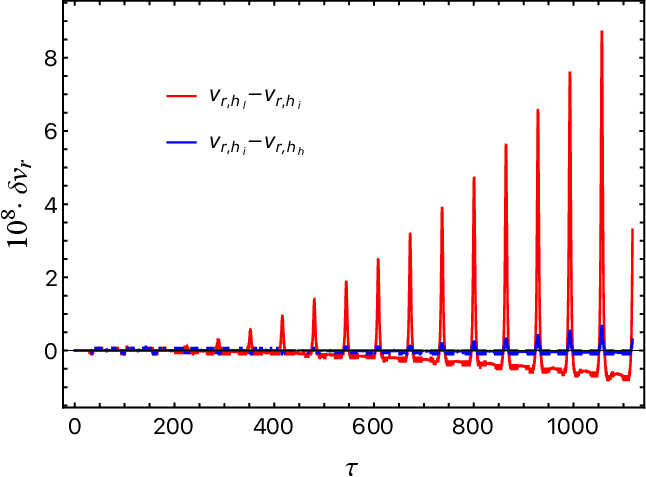}
\caption{Convergence test for the radial velocity $v_r$. The red and blue curves represent the residuals between coarse-intermediate $(v_{r,h_l}-v_{r,h_i})$ and intermediate-high $(v_{r,h_i}-v_{r,h_h})$ resolutions, respectively.}
\label{Convergence_Orbits}
\end{center}
\end{figure}

To guarantee the robustness and accuracy of the orbital integration, we perform a self-convergence test for the radial velocity $v_r$. We employ three distinct temporal resolutions by successively halving the proper time step: coarse ($h_l: \delta\tau = 0.03$), intermediate ($h_i: \delta\tau = 0.02$), and high ($h_h: \delta\tau = 0.01$). Fig.~\ref{Convergence_Orbits} illustrates the evolution of the residuals between these successive resolutions over the simulation duration. As the step size is refined, the residuals decrease systematically, yielding a calculated convergence factor of $Q = |v_{r,h_l} - v_{r,h_i}| / |v_{r,h_i} - v_{r,h_h}| \approx 13$. This result is in good agreement with the theoretical value of $16$ expected for a fourth-order Runge-Kutta (RK4) scheme, confirming that the numerical integration is well-converged and stable.

\begin{figure*}[t]
\begin{center}
\subfigure[~$\ell=-0.01$]  {\label{Fig_Orbit_l_m001_L3E0985}
\includegraphics[width=4.6cm]{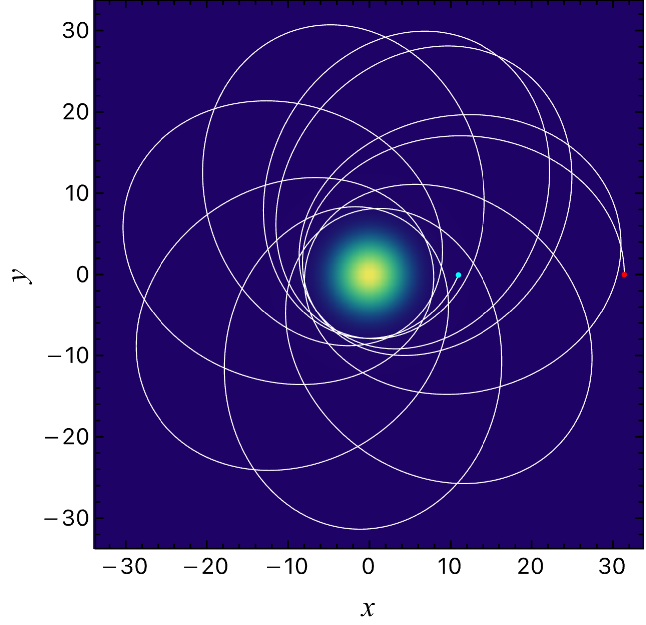}}
\subfigure[~$\ell=0$]  {\label{Fig_Orbit_l_0_L3E0985}
\includegraphics[width=4.6cm]{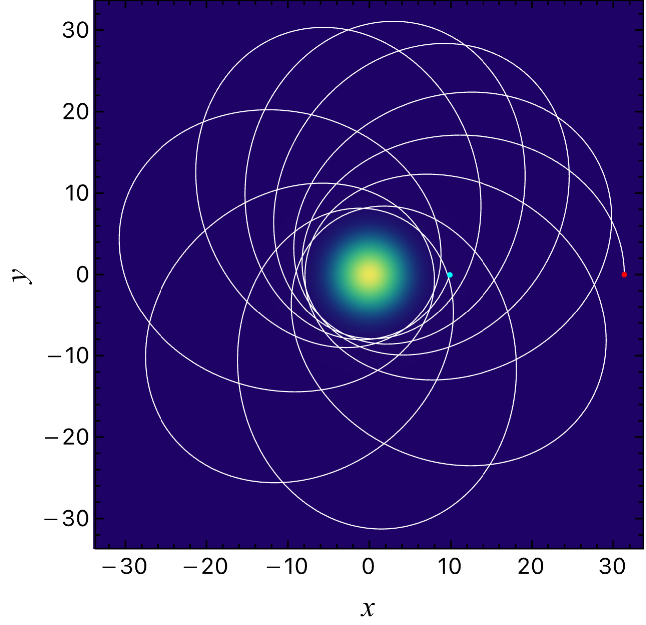}}
\subfigure[~$\ell=0.01$]  {\label{Fig_Orbit_l_p001_L3E0985}
\includegraphics[width=5.5cm]{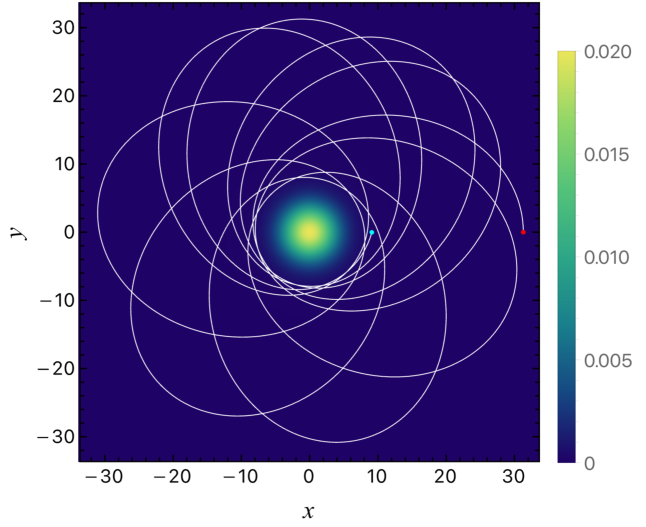}}
\caption{Orbital trajectories for different values of the Lorentz-violating parameter $\ell$, computed for $(L=3, E=0.985)$ and $\omega=0.9$. The coordinate range is $x, y \in [-32, 32]$. The background green density plot shows the distribution of the scalar field's energy density.}
\label{Fig_Orbit_L3E0985}
\vspace{10pt}
\subfigure[~$\ell=-0.01$]  {\label{Fig_Orbit_l_m001_L05E09}
\includegraphics[width=4.6cm]{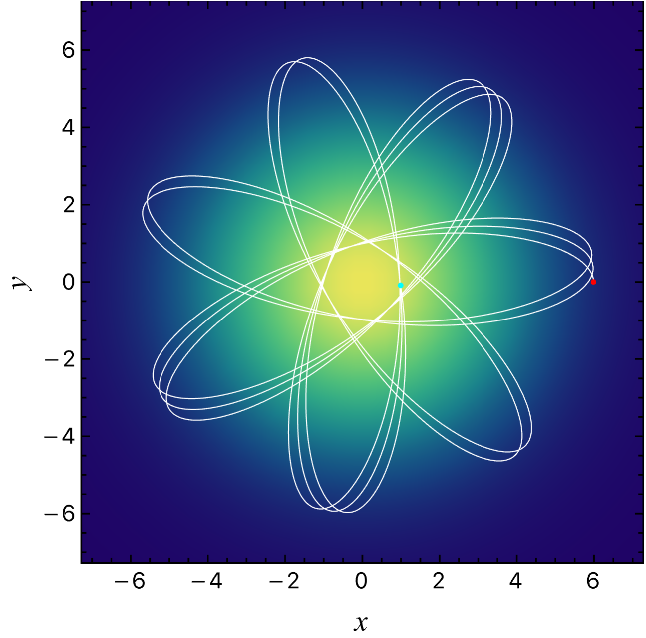}}
\subfigure[~$\ell=0$]  {\label{Fig_Orbit_l_0_L05E09}
\includegraphics[width=4.6cm]{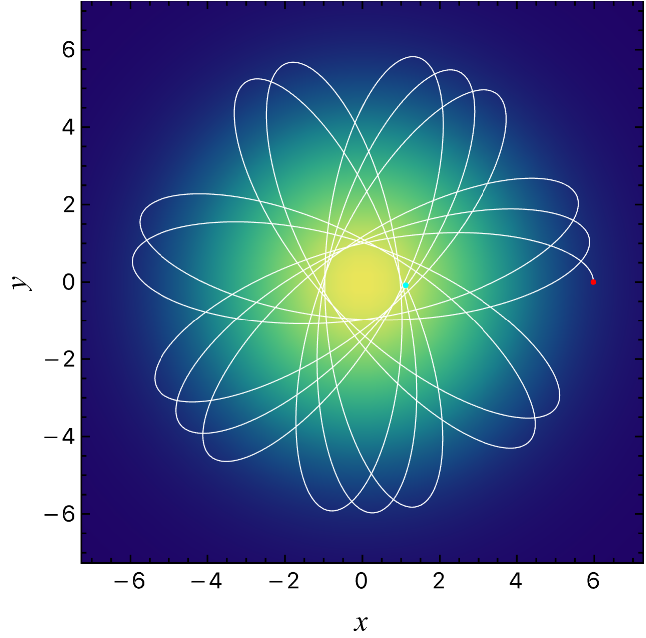}}
\subfigure[~$\ell=0.01$]  {\label{Fig_Orbit_l_p001_L05E09}
\includegraphics[width=5.5cm]{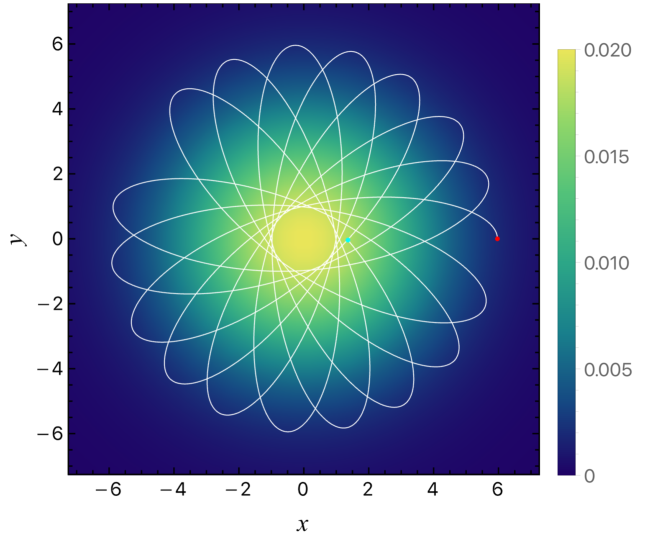}}
\caption{Orbital trajectories for different values of the Lorentz-violating parameter $\ell$, computed for $(L=0.5, E=0.9)$ and $\omega=0.9$. Note that this figure shows a zoomed-in view of the central region with a smaller coordinate range $x, y \in [-7, 7]$. The background green density plot shows the distribution of the scalar field's energy density.}
\label{Fig_Orbit_L05E095}
\end{center}
\end{figure*}

The orbital trajectories associated with the effective potentials in Figs.~\ref{Fig_Potential_L3E0985} and \ref{Fig_Potential_L05E09} are plotted in Figs. \ref{Fig_Orbit_L3E0985} and \ref{Fig_Orbit_L05E095}, respectively. In each panel, the scalar‑field energy density is indicated by the green colormap. The trajectories display a ``rosette" pattern, typical of bound eccentric orbits in a non-Newtonian potential. The fact that the loops do not close reveals a significant periastron precession.

As shown in Fig.~\ref{Fig_Orbit_L3E0985}, a larger angular momentum ($L = 3$) produces a stronger centrifugal barrier, confining the test particle to grazing orbits with a larger average radius outside the boson star's core, spanning a coordinate range of $x, y \in [-32, 32]$. In contrast, as illustrated in Fig.~\ref{Fig_Orbit_L05E095}, a smaller angular momentum ($L = 0.5$) yields a weaker barrier, resulting in more compact and higher‑eccentricity penetrating orbits that allow the particle to penetrate deeply into the core, with the coordinate range restricted to $x, y \in [-7, 7]$. Furthermore, variations in the Lorentz-violating parameter $\ell$ induce a noticeable shift in the orbital precession angle.

The orbital shape can be quantified using the eccentricity $\varepsilon$, defined as 
\bal
\varepsilon=\fc{r_a-r_p}{r_a + r_p},
\label{eccentricity}
\eal
where $r_a$ and $r_p$ denote the apastron and periastron distances of the orbit, respectively. The corresponding orbital parameters for different $\ell$ values are tabulated in Tabs.~\ref{Tab_Orbits_L3E0985} and~\ref{Tab_Orbits_L05E09}. As shown in these tables, the eccentricity decreases as the values of $\ell$ increase from negative to positive. This implies that a positive $\ell$ favors more circularized orbits than the GR case ($\ell=0$), whereas a negative $\ell$ results in more eccentric trajectories. This observed trend can be intuitively understood through the impact of $\ell$ on the radial pressure $P_r$, which provides the hydrodynamic support against gravity. 

As $\ell$ increases from negative to positive values, the radial pressure $P_r$ is progressively suppressed, thereby enhancing the gravitational binding of the boson star and yielding a more compact central configuration. This structural evolution yields a deeper and narrower effective potential well $V_{\text{eff}}$, as illustrated in Fig.~\ref{Fig_Potentials}, which directly governs the orbital turning points. Specifically, the apastron $r_a$ decreases while the periastron $r_p$ increases, causing the radial range $\Delta r = r_a - r_p$ to contract, a trend clearly evident in Tabs.~\ref{Tab_Orbits_L3E0985} and~\ref{Tab_Orbits_L05E09}. In the eccentricity formula \eqref{eccentricity}, the numerator $\Delta r$ diminishes rapidly, whereas the denominator $r_a + r_p$ remains relatively stable due to the partial cancellation between the inward shift of $r_a$ and the outward shift of $r_p$. Thus, the contraction of the radial range dominates the ratio, resulting in the monotonic decrease in eccentricity $\varepsilon$ observed in the data.

\begin{table}[t]
\begin{center}
\begin{tabular}{|c|c|c|c|c|c|}
\hline
$\ell$  & $r_p/M$  &  $r_a/M$ &  $\Delta r/M$   & $\varepsilon$  & $T_{2\pi}$    \\
\hline
-0.01   & 7.8965 & 31.3920  & 23.4955 & 0.5980  & 3256.45    \\
\hline
0  & 7.9141 & 31.3447  & 23.4305 & 0.5968  & 3264.74    \\
\hline
0.01   & 7.9323 & 31.2942 &  23.3619 & 0.5956  & 3273.03    \\ 
\hline
\end{tabular}\\
\caption{Periastron $r_p$, apastron $r_a$, radial range $\Delta r$, eccentricity $\varepsilon$, and azimuthal orbital period $T_{2\pi}$ associated with the orbits in Fig.~\ref{Fig_Orbit_L3E0985}.}
\label{Tab_Orbits_L3E0985}
\vspace{10pt}
\begin{tabular}{|c|c|c|c|c|c|}
\hline
$\ell$  & $r_p/M$  &  $r_a/M$ &  $\Delta r/M$   & $\varepsilon$  & $T_{2\pi}$    \\
\hline
-0.01   & 0.9672& 5.9871  & 5.0200 & 0.7218  & 546.37    \\
\hline
0  & 0.9721 & 5.9765  & 5.0045 & 0.7202  & 548.88   \\
\hline
0.01   & 0.9769 & 5.9658 &  4.9889 & 0.7186  & 551.37    \\
\hline
\end{tabular}\\
\caption{Periastron $r_p$, apastron $r_a$, radial range $\Delta r$, eccentricity $\varepsilon$, and azimuthal orbital period $T_{2\pi}$ associated with the orbits in Fig.~\ref{Fig_Orbit_L05E095}.}
\label{Tab_Orbits_L05E09}
\end{center}
\end{table}

\section{Gravitational Waves from EMRIs }\label{Gravitational_Waves}

EMRI systems are among the most promising sources for low-frequency, space-based gravitational-wave observatories. Over many orbital cycles, the gravitational-wave signal from an EMRI gradually accumulates subtle deviations from the predictions of GR. This cumulative effect renders EMRIs highly sensitive probes for testing GR and detecting potential beyond-GR signatures.

In this section, we investigate the gravitational waveforms arising from the orbits delineated in the previous section, with the aim of elucidating the effects of Lorentz violation on these signals. In fact, these orbits are inherently dynamic, as GW emission drives a gradual dissipation of energy and angular momentum, thereby inducing an inspiral that alters the orbital parameters over time. However, our analysis focuses on a brief segment spanning only a few orbital cycles, during which the radiation-reaction effects remain negligible relative to the orbital timescale. Consequently, the adiabatic approximation holds robustly, allowing us to model the evolution as quasi-static. To generate the corresponding EMRI gravitational waveforms, we employ the numerical Kludge method \cite{Babak2007}, which has been widely used in literature \cite{Junior2025,Yang2025}.

In the preceding section, the orbital trajectories were derived within Schwarzschild coordinates. Following the methodology outlined in Ref.~\cite{Babak2007}, the ``equivalent'' flat-space trajectories can be constructed by projecting these coordinates onto a flat-space Cartesian system via the transformation
\bal
x=r \sin\theta \cos\varphi, \quad y=r \sin\theta \sin\varphi, \quad z=r \cos\theta.
\eal
These flat-space trajectories provide the basis for generating gravitational waveforms. 

The gravitational-wave strain can be approximated using the quadrupole formula \cite{Thorne1980}:
\bal
h_{ij}=\lt.\fc{2}{D_\tx{L}}{\fc{d^2{I}_{ij}(t')}{dt'{}^2}}\rt|_{t'=t-D_\tx{L}},
\label{Quadrupole_Formula}
\eal
where $D_\tx{L}$ denotes the luminosity distance from the EMRI system to the detector, and the symmetric trace-free mass quadrupole moment is defined as 
\bal
I^{ij}(t')=\int \rho(t',\bm{x'})\lt(x'{}^i x'{}^j-\fc{1}{3}\delta^{ij}r'{}^2\rt) d^3x'.
\label{Quadrupole}
\eal

Under the point-mass approximation, the mass density of the orbiting body on a trajectory $\bm{Z}(t)$ is given by
\bal
\rho(t,\bm{x})=m_\tx{s}\delta^3\lt(\bm{x}-\bm{Z}(t) \rt), 
\eal
with $m_\tx{s}$ its mass. 

After evaluating the integrals and derivatives, the explicit expression for the gravitational-wave strain \eqref{Quadrupole_Formula} becomes \cite{Yang2025}
\bal
h_{ij}=\fc{2m_\tx{s}}{D_\tx{L}}\lt[a_i x_j + a_j x_i + 2v_i v_j -\fc{2}{3}\delta_{ij}\lt(\bm{a}\cdot\bm{x} + v^2\rt)\rt],
\label{Gravitational_Radiation}
\eal
where $a_i$ and $v_i$ are the Cartesian components of the orbiting body’s acceleration and velocity, respectively. 

To construct the observable polarizations, it is useful to introduce a detector-adapted coordinate system $(X, Y, Z)$ centered on the supermassive object. The basis vectors of this frame are expressed in the original coordinates as \cite{Poisson2014}
\bal
\bm{e}_X &= [\cos\zeta, \, -\sin\zeta, \, 0],\\
\bm{e}_Y &= [\cos\iota\sin\zeta, \, \cos\iota\cos\zeta, \, -\sin\iota],\\
\bm{e}_Z &= [\sin\iota\sin\zeta, \, \sin\iota\cos\zeta, \, \cos\iota],
\eal
where $\iota$ is the inclination angle of the orbital plane with respect to the $X$-$Y$ plane, and $\zeta$ is the longitude of the pericenter measured within the orbital plane.

\begin{figure*}[t]
\begin{center}
\subfigure[~$h_+$]  {\label{Fig_GW_l_001_plus_L3E0985}
\includegraphics[width=13cm]{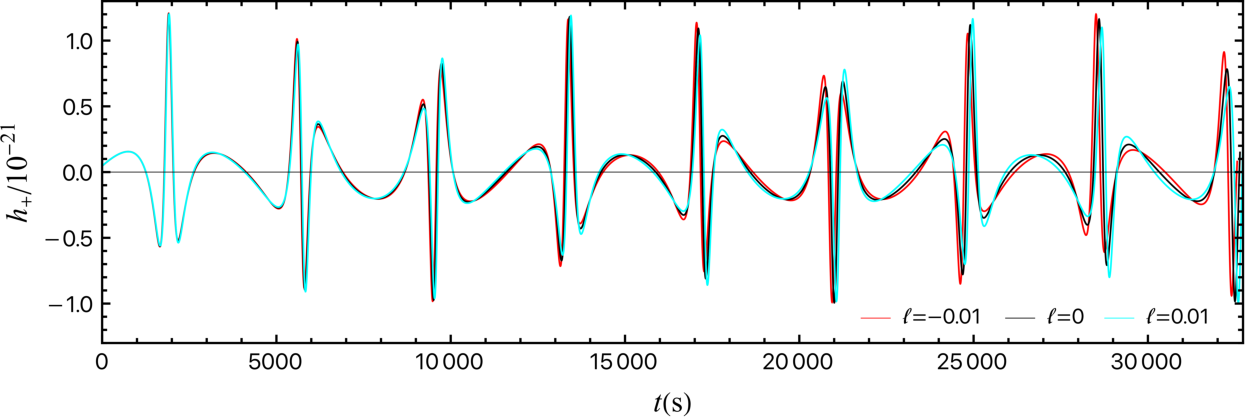}}
\subfigure[~$h_\times$]  {\label{Fig_GW_l_001_cross_L3E0985}
\includegraphics[width=13cm]{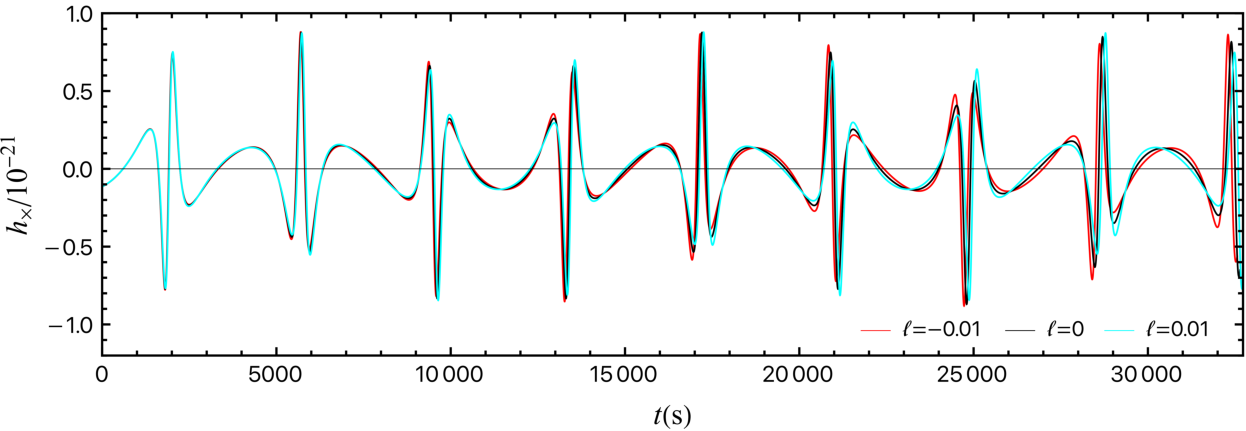}}
\caption{Gravitational waveforms for different values of the Lorentz-violating parameter $\ell$ corresponding to trajectories with $(L=3, E=0.985)$, where $\omega=0.9$.}
\label{Gravitational_Waveforms_L3E0985}
\end{center}
\end{figure*}

\begin{figure*}[t]
\begin{center}
\subfigure[~$h_+$]  {\label{Fig_GW_l_001_plus_L05E09}
\includegraphics[width=13cm]{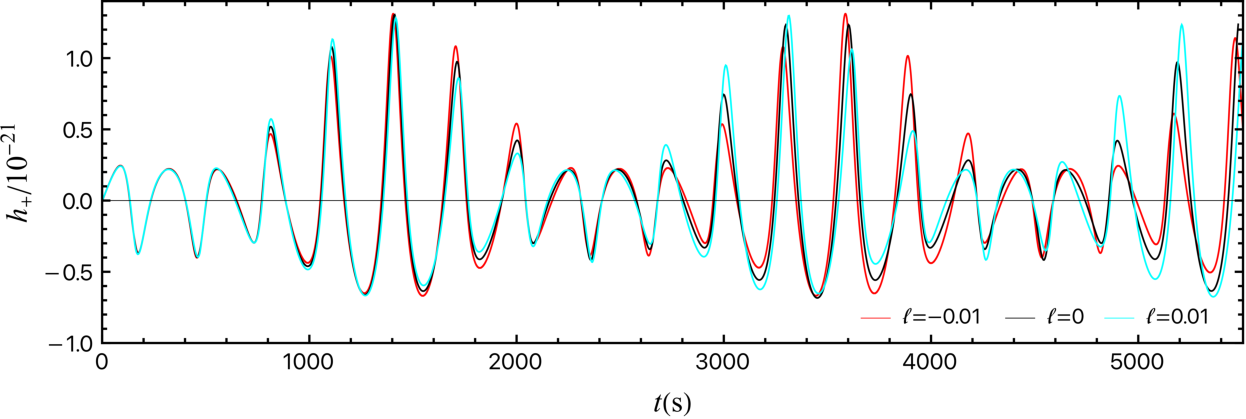}}
\subfigure[~$h_\times$]  {\label{Fig_GW_l_001_cross_L05E09}
\includegraphics[width=13cm]{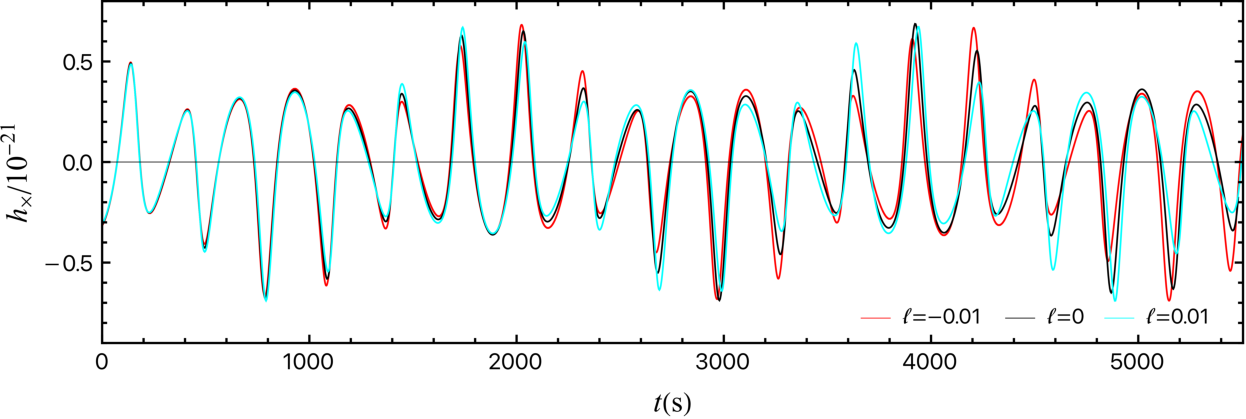}}
\caption{Gravitational waveforms for different values of the Lorentz-violating parameter $\ell$ corresponding to trajectories with $(L=0.5, E=0.9)$, where $\omega=0.9$.}
\label{Gravitational_Waveforms_L05E09}
\end{center}
\end{figure*}

The physical polarizations $h_+$ and $h_\times$ are obtained by projecting the tensor $h_{ij}$ onto these basis vectors, yielding \cite{Babak2007}
\bal
h_+ &= \fc{1}{2}\lt(\bm{e}_X^i \bm{e}_X^j - \bm{e}_Y^i \bm{e}_Y^j\rt) h_{ij}=\fc{1}{2}\lt(h_{\zeta\zeta} - h_{\iota\iota}\rt),\\
h_\times &= \fc{1}{2}\lt(\bm{e}_X^i \bm{e}_Y^j + \bm{e}_Y^i\bm{e}_X^j\rt) h_{ij}=h_{\iota\zeta},
\eal
where the projected components $h_{\zeta\zeta}$, $h_{\iota\iota}$, and $h_{\iota\zeta}$ are given by 
\bal
h_{\zeta\zeta} =& h_{xx} \cos^2\zeta - h_{xy} \sin 2\zeta + h_{yy} \sin^2\zeta,\\
h_{\iota\iota} =& \cos^2\iota \lt( h_{xx} \sin^2\zeta+  h_{xy} \sin 2\zeta + h_{yy} \cos^2\zeta\rt)+h_{zz}\sin^2\iota-\sin 2\iota \lt(h_{xz}\sin\zeta+h_{yz}\cos\zeta \rt),\\
h_{\iota\zeta} =& \cos\iota \lt(\fc{1}{2}h_{xx}\sin2\zeta+h_{xy}\cos2\zeta-\fc{1}{2}h_{yy}\sin2\zeta \rt)+\sin\iota \lt(h_{yz} \sin\zeta-h_{xz}\cos\zeta \rt).
\eal
Note that although Eq.~\eqref{Gravitational_Radiation} is already trace-free, the projection operation automatically filters out the trace part and extracts the relevant radiative components.

\begin{figure*}[t]
\begin{center}
\subfigure[~$(L=3, E=0.985)$]  {\label{Fig_hc_L3E0985}
\includegraphics[width=7cm,height=5cm]{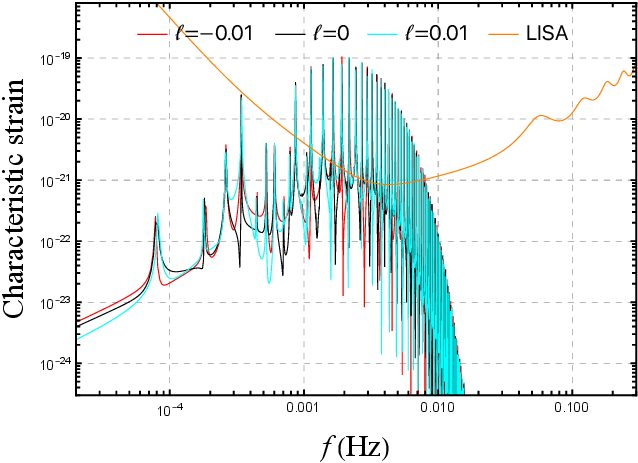}}\qquad
\subfigure[~$(L=0.5, E=0.9)$]  {\label{Fig_hc_L05E09}
\includegraphics[width=7cm,height=5cm]{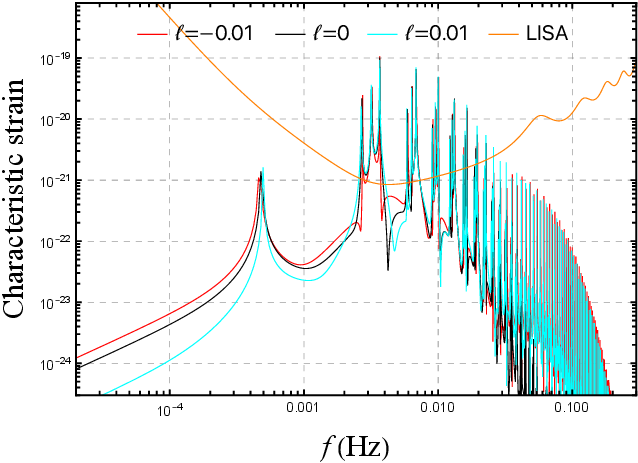}}
\caption{Characteristic strain for different values of the Lorentz-violating parameter $\ell$. Panels (a) and (b) correspond to waveforms with $(L=3, E=0.985)$ and $(L=0.5, E=0.9)$, respectively. The LISA sensitivity curve is shown in orange for comparison.}
\label{Characteristic_strain}
\end{center}
\end{figure*}

We model an EMRI system in which a stellar-mass compact object of mass $m_\tx{s} = 10 M_\odot$ orbits a central boson star of mass $M = 10^6 M_\odot$. The source is located at a luminosity distance of $D_\tx{L} = 0.1\,$Gpc, with an inclination angle $\iota = \pi/3$ and a longitude of pericenter $\zeta = \pi/3$. The corresponding gravitational waveforms for the orbital configurations illustrated in Figs.~\ref{Fig_Orbit_L3E0985} and \ref{Fig_Orbit_L05E095} are shown in Figs.~\ref{Gravitational_Waveforms_L3E0985} and \ref{Gravitational_Waveforms_L05E09}, respectively.

As shown in Fig.~\ref{Gravitational_Waveforms_L3E0985}, grazing orbits produce intermittent gravitational-wave bursts separated by long quiescent intervals, resembling the signal morphology of a black-hole EMRI. This resemblance is apparent in the waveform structure: a smooth inward drift from apastron culminates in a sharp radiation burst during the sudden ``whiplash” turnaround near periastron, generating distinct, intermittent spikes in the strain.

In contrast, penetrating orbits traverse regions of high scalar‑field density within the boson star. The effective‑potential analysis in Fig.~\ref{Fig_Potentials} shows that these orbits are confined within a significantly deeper and narrower potential well compared to the $L=3$ case. Consequently, as illustrated in Fig.~\ref{Gravitational_Waveforms_L05E09}, the particle remains in a state of sustained high acceleration, emitting sustained, modulated gravitational-wave oscillations with no quiescent phase. This behavior reflects the unique signature of a particle bound within the high‑density scalar‑field core.

At the early stages of the inspiral, the waveforms incorporating Lorentz violation show close agreement with the GR case. However, as the orbital evolution progresses, the Lorentz-violating parameter $\ell$ induces a cumulative phase shift, leading to a progressive de-phasing of the gravitational-wave signals. To quantitatively characterize this divergence, we define the normalized strain deviation as
\bal
    \Delta h(t) = \fc{|h_{\ell}(t) - h_{\tx{GR}}(t)|}{h_{\tx{peak}}},
\eal
where $h_{\ell}(t)$ represents the strain amplitude of either the plus or cross polarization for the Lorentz-violating waveform, and $h_{\text{peak}}$ is the peak strain amplitude of the corresponding GR waveform. By adopting a threshold of $\Delta h = 10\%$ to define the onset of discernible de-phasing, we find that the signals reach this criterion at markedly different timescales. 

Specifically, for the grazing orbits ($L=3, E=0.985$), the threshold is reached at $t = 5647$ s for the plus polarization (Fig.~\ref{Fig_GW_l_001_plus_L3E0985}) and $t = 5562$ s for the cross polarization (Fig.~\ref{Fig_GW_l_001_cross_L3E0985}). In contrast, for the penetrating orbits ($L=0.5, E=0.9$), the $10\%$ deviation occurs much earlier, occurring at $t = 1359$ s for the plus polarization (Fig.~\ref{Fig_GW_l_001_plus_L05E09}) and $t = 1694$ s for the cross polarization (Fig.~\ref{Fig_GW_l_001_cross_L05E09}). This indicates that penetrating orbits serve as significantly more sensitive probes of Lorentz-violating effects.

To quantify the detection prospects for gravitational waves emitted by the EMRIs, we compute the characteristic strain $h_c(f)$ given by \cite{Robson2019}
\bal
h_c(f)=2f\sqrt{\lt|\tilde{h}_+(f)\rt|^2+\lt|\tilde{h}_\times(f)\rt|^2 },
\eal
where $\tilde{h}_{+}(f)$ and $\tilde{h}_{\times}(f)$ denote the Fourier transforms of the plus and cross polarizations, respectively. The results are compared with the LISA sensitivity curve \cite{Robson2019} in Fig.~\ref{Characteristic_strain}. 

As shown in Fig.~\ref{Fig_hc_L3E0985}, grazing orbits emit most of their gravitational-wave power below $0.01\,$Hz. In contrast, Fig.~\ref{Fig_hc_L05E09} reveals that penetrating orbits exhibit a markedly richer spectrum extending beyond $0.1\,$Hz.  This spectral enhancement stems from the particle's rapid periodic motion within the deep effective potential well. Indeed, the time-domain waveforms reveal a significantly shorter orbital period for penetrating orbits ($T \approx 550\,$s compared to $T \approx 3200\,$s for grazing orbits), which directly establishes a higher fundamental frequency. In both cases, the signal peaks exceed the LISA sensitivity threshold within its most sensitive frequency band, indicating their potential detectability with future space-based observatories.

The Lorentz-violation effect induces discernible frequency shifts and intensity modulations in the harmonic peaks, particularly at low frequencies. A comparison of detectability further reveals a key distinction between the two orbital regimes. For grazing orbits (Fig.~\ref{Fig_hc_L3E0985}), the harmonic peaks that surpass the LISA sensitivity threshold across different $\ell$ values remain closely clustered and nearly indistinguishable. In contrast, penetrating orbits (Fig.~\ref{Fig_hc_L05E09}) exhibit apparent frequency shifts and intensity modulations in peaks around $f \sim 0.003\,$Hz, all robustly exceeding the LISA sensitivity curve. Consequently, these distinctive modulation features in penetrating orbits could therefore provide a potential observational signature for constraining Lorentz violation with future space-based gravitational-wave detectors.

\section{Conclusions}\label{Conclusions}

In this work, we investigated the impact of spontaneous Lorentz symmetry breaking on the mini-boson stars within the framework of bumblebee gravity. We considered spherically symmetric, static equilibrium configurations and numerically solved the resulting modified field equations. Our analysis reveals that Lorentz violation induces clear deviations in the scalar profile $\phi(r)$, the mass function $m(r)$, and the metric functions $n(r)$ and $\sigma(r)$ from the standard GR case.

Our numerical results indicate that the Lorentz-violating parameter $\ell$ effectively influences boson-star cohesion by modulating the radial pressure $P_r$. In particular, for $\ell > 0$, the suppression of the repulsive pressure manifests as an effective attractive interaction, thereby strengthening gravitational binding. This leads to stronger radial confinement, an increase in the boson-star mass $M$, particle number $N$, and compactness $C$, as well as a broadening of the solution space. In contrast, for $\ell < 0$, the enhanced repulsive pressure introduces an effective repulsion that promotes outward diffusion. As $\ell$ decreases continuously from positive to negative values, this growing repulsion progressively weakens gravitational binding until, below a critical negative threshold, static boson-star configurations can no longer be supported.

These structural modifications leave distinct imprints on the orbital dynamics and gravitational-wave emission of EMRIs around boson stars. A positive Lorentz-violating parameter $\ell$ yields more compact, more circular orbits with reduced eccentricity and smaller radial range, whereas a negative $\ell$ leads to more extended and more eccentric trajectories. 

Using the numerical Kludge method, we computed the corresponding gravitational waveforms. We found that grazing orbits, which are confined to larger radii, produce intermittent gravitational-wave bursts separated by long quiescent intervals, closely resembling those of black‑hole EMRIs. In contrast, penetrating orbits that enter the boson‑star core generate sustained, modulated oscillations without a quiescent phase. As the inspiral evolves, waveforms corresponding to different $\ell$ values exhibit progressively accumulating phase and amplitude differences. 

To assess the detectability of the emitted gravitational waves, we also computed the characteristic strain and compared it with the LISA sensitivity curve. In contrast to the grazing orbits, where harmonic peaks remain clustered and nearly indistinguishable, the penetrating orbits exhibit apparent frequency shifts and intensity modulations that rise above the LISA sensitivity curve, offering a potential observable for constraining Lorentz violation with future space‑based detectors.

In the EMRI calculations, we adopted Lorentz‑violating parameters $\ell=\pm 0.01$ as an illustrative example to analyze their dynamical effects and evaluate their imprint on the gravitational-wave signals. Given that existing experimental bounds are far more stringent than the parameter values adopted here, such as $|\ell| < 10^{-10}$ from solar system tests \cite{Casana2018}, detecting such subtle Lorentz-violating effects would require simulations spanning longer orbital evolution intervals, fully incorporating dissipative mechanisms including gravitational radiation reaction and dynamical friction. Such extended studies lie beyond the scope of this paper and remain for future investigation.

\section*{ACKNOWLEDGMENTS}

We thank the referee for the constructive comments and suggestions, which have significantly improved the quality of this manuscript. We are also grateful to Yu-Peng Zhang, Sen Yang, Shou-Long Li and Yu-Meng Xu for helpful discussions. This work was supported by the National Natural Science Foundation of China (Grant No.~12475062) and the Natural Science Foundation of Chongqing (Grant No.~CSTB2024NSCQ-MSX0358). KY also acknowledges the generous hospitality during a visit to the Lanzhou Center for Theoretical Physics, where part of this work was completed.


\end{document}